\documentstyle[12pt]{article}
\newcommand\be{\begin{equation}}
\newcommand\ee{\end{equation}}
\newcommand\bea{\begin{eqnarray}}
\newcommand\eea{\end{eqnarray}}

\newcommand{\fatalpha}{{\bf \alpha \kern -0.44em \alpha}}
\newcommand{\fatsigma}{{\bf \sigma \kern -0.54em \sigma}}
\newcommand{\tpchi}{{\bf \chi \kern -0.35em \chi}}
\newcommand{\llambda}{{\bf \lambda \kern -0.45em \lambda}}

\renewcommand{\theequation}{\arabic{equation}}
\renewcommand{\theequation}{\thesection-\arabic{equation}}
\bibliography{plain}
\title{\bf Asymptotic entanglement in 1D quantum walks with a time-dependent coined }\vspace{20mm}
\author{ S. Salimi \thanks{Corresponding author:  E-mail addresses:
  shsalimi@uok.ac.ir},
   R. Yosefjani
  \thanks{E-mail addresses: R.yousefjany@uok.ac.ir}
 \\ {\small Department of Physics,
University of Kurdistan, P.O.Box 66177-15175 , Sanandaj, Iran.}}
\pagebreak
\vspace{20mm}
\begin{document}
\maketitle \vspace{10mm}
\begin{abstract}
Discrete-time quantum walk evolve by a unitary operator which
involves two operators a conditional shift in position space and a
coin operator. This operator entangles the coin and position degrees
of freedom of the walker. In this paper, we investigate the
asymptotic behavior of the coin position entanglement (CPE) for an
inhomogeneous quantum walk which determined by two orthogonal
matrices in one-dimensional lattice. Free parameters of coin
operator together provide many conditions under which a measurement
perform on the coin state yield the value of entanglement on the
resulting position quantum state. We study the problem analytically
for all values that two free parameters of coin operator can take
and the conditions under which entanglement becomes maximal are
sought.
\end{abstract}

\section{Introduction}
Among classical algorithms, many are based on classical random walk. Markov chain simulation, which has emerged as
a powerful algorithmic tool \cite{1} is one such
example. Like classical random walk, the quantum version of it has also become
an important constituent of quantum algorithms and computation.
Though quantum random walk was first introduced by Aharonov et al. \cite{2},
but unlike its classical counterpart, the evolution of the quantum version is
unitary, reversible and has no randomness associated with it during the evolution.
Therefore, keeping away the term 'random', quantum walk has been the preferred usage.
In the classical random walk the particle moves in the position space with a certain probability,
whereas the quantum walk, which involves a superposition of states, moves by exploring multiple possible paths
 simultaneously with the amplitudes corresponding to different paths interfering. This properties makes
 the variance of the quantum walk on a line to grow quadratically with the number of steps (time),
 compared to the linear growth for the classical random walk \cite{3,4}. Quantum walk is studied in
 two standard forms: continuous-time quantum walk and discrete-time quantum walk that was introduced
 by Farhi and Gutmann \cite{5} and Watrous \cite{6}, respectively. In the continuous-time quantum walk,
 one can directly define the walk on the position Hilbert space \cite{5},  but in the discrete-time
 quantum walk, in addition to it is necessary to introduce  a coin Hilbert space to define the direction in
 which the particle has to move \cite{3}. Due to the coin degree of freedom, the discrete-time variant is
 shown to be more powerful than the other in some contexts \cite{7}. Childs describe a precise correspondence
 between continuous- and discrete-time quantum walks on arbitrary graphs \cite{8}. Both these types of quantum walk have been
widely used in algorithms for a diverse of problems. See for example \cite{9,10,11,12,13,14}.
Beyond applications in quantum algorithms, quantum walk is emerging as a potential tool to understand various
phenomenon in physical systems and has been
employed to demonstrate coherent control over quantum many body systems. See for example \cite{15,16,17,18}.
Some experimental progress on the implementation of quantum walk has been reported \cite{19,20,21,22,23,24}.
\\ Entanglement is one of the attractive properties of quantum that does not appear in classic and is a very
useful resource to perform various quantum tasks, a recent review by Horodecki et al. \cite{25} discuss many of
these aspects.
Unfortunately, physical limitations and noise effects change the amount of entanglement and restrict it's efficiency.
One way of circumventing this problem would be to generate entanglement. Quantum walk is one such process in which the conditional
 shift operator, which governs the itinerary of the quantum walker, induces entanglement between the degree of freedom of the coin
 and the spatial degree of freedom of the walker. This entanglement fluctuates with each step and eventually settles down to an
 asymptotic value that depends on the initial state of the quantum walk and on any bias in the quantum coin operator. Entanglement
  is a basic resource in quantum algorithms, and further work is required in order to fully understand its properties in the QW.
While, entanglement between coin and position of the walker, as a type of quantum correlation, distort the
distribution of the quantum walk and give it peaks and troughs, especially at the ends of the top hat \cite{52,53}, it is necessary
to obtain linear spreading and mixing times \cite{54} and can be use to gauge the impact of the added randomness and
 decoherence \cite{51}. In addition, it can be exploited for quantum information and communication purposes \cite{47,48,49,50}.
Entanglement has also given us new insights for understanding many physical phenomena including super-radiance \cite{26},
 superconductivity \cite{27}, disordered systems \cite{28} and
emerging of classicality \cite{29}. In particular, understanding the role of entanglement in the existing methods of simulations
of quantum spin
systems allowed for significant improvement of the methods, as well as understanding their limitations \cite{30}. So,
 studying entanglement during the quantum walk process will be useful from a quantum information theory perspective and
 also contribute to further investigation of the
practical applications of the quantum walk.
Carneiro et al. \cite{33} studied the long-time asymptotic coin-position entanglemant of quantum walks on various
graphs ($Z$ , $Z_{2}$ , triangular lattices, cycles). Venegas-Andraca and Bose \cite{34} also investigated generation
of entanglement between two walkers. Using Fourier analysis techniques, Abal et al. \cite{35} analytically computed
asymptotic coin position entanglement of the Hadamard walk on one dimension for both localized and non-localized initial
conditions and in the same way Annabestani et al. \cite{36} studied asymptotic entanglement in a two-dimensional quantum walk.
Chandrashekar et al. \cite{37} consider a multipartite quantum walk on a one-dimensional lattice and studeid the evolution of
spatial entanglement, entanglement between different lattice points.
\\Most of the original work considered homogeneous walks, which the amplitude for moving does not depend on position. The idea
of looking at inhomogeneous walks is not new, the recognition that
it is natural to allow coins to be position dependent may be found
in \cite{38,39,40,41,42,55}. To probe the possible long time
behaviours and quantify the entanglement between the coin and
position of such walks is our motivation for this work. In this
paper, we investigate the asymptotic behavior of the  CPE  for an
inhomogeneous QW which is determined by two orthogonal matrices in
one-dimensional lattice. Our inhomogeneous QW is a kind of the
generalized model defined in \cite{55, 56} which is based on the
idea of the Aubry-Andr\'{e} model \cite{57}. The limit distribution
of the this QW has the probability density given by  Dirac delta
function  which is called a localization for the inhomogeneous QW
(for more details see Ref.\cite{55}).  We show how the entanglement
approaches asymptotic values depending on the choice of initial
state and coin bias with study the problem analytically for all
values of $\theta_{0}$ and $\theta_{1}$ for large number of QW
steps. We find that CPE is different for odd or even positions and
when on the coin parameters took the precise value $\frac{\pi}{2}$
walk alter to a bounded motion.
\\ The plan of the paper is as follows: in section $2$
the model for discrete-time quantum walk is presented in detail.
Section $3$ defines the entanglement and provides the required formalism in
the Fourier space leading to the long-time reduced density operator for arbitrary
coin operations and initial states. Our inhomogeneous walk with two-period has defined
in section $4$ and the asymptotic CPE entanglement of our walk as a function of coin operators
for local and non-local initial conditions is obtained. Conclusions are given in the last part, section $5$.

\section{Definition of the quantum walk}
The discrete-time quantum walk (DTQW) on a line is defined on a Hilbert space $\mathbf{H} = \mathbf{H}_{p}\otimes \mathbf{H}_{c}$,
 where $\mathbf{H}_{c}$ is the coin Hilbert space, spanned by the basis state $|L\rangle = [1,0]^{T}$ and $|R\rangle = [0,1]^{T}$,
 where $T$ denotes the transposed operator, that represents two sides of the directions of the motion and $\mathbf{H}_{p}$ is the
 position Hilbert space, spanned by the infinite basis states $\{|x\rangle ; x\in \textbf{Z}\} $ that represent
 the position of the walker. In one step of the QW we first make superposition on the coin space with a coin
 operator $C\in U(2)$ and after that we move the particle according to the coin state with the translation
 operator $S$ as $U = S .(I_{p} \otimes C)$, where $S$ is defined as
 $S = \Sigma_{x} |x-1\rangle\langle x| \otimes |L\rangle\langle L| + |x+1\rangle\langle x| \otimes |R\rangle\langle R|$, and $I_{p}$
 is the identity operator in $\mathbf{H}_{p}$. The most widely studied form of the DTQW is the walk
 using the Hadamard operation as quantum coin. It is an unbiased coin operation, the resulting walk is known as the
  Hadamard walk, which determined by the Hadamard gate: $H = \frac{1}{\sqrt{2}} \left(\matrix{1&1 \cr 1&-1}\right)$.
\\ The state of QW at time $t$ is expressed by
\begin{equation}\label{2-1}
|\Psi(t)\rangle=\sum_{x}|x\rangle \otimes|\psi(x,t)\rangle,
\end{equation}
where $|\psi(x,t)\rangle=a(x,t)|L\rangle + b(x,t)|R\rangle$ denotes the coin state and $a(x,t)$ $(b(x,t))$ is the amplitude of the base $|x,L\rangle$ $(|x,R\rangle)$ at time $t$ which belong to the complex number $\textbf{C}$, satisfying the normalization condition $\sum_{x}|a(x,t)|^{2}+|b(x,t)|^{2}=1$. Total state after one steps, is given by $|\Psi(t+1)\rangle = U|\Psi(t)\rangle$. The Fourier transform, as first noted in this context by Nayak and Vishwanath \cite{4}, is extremely useful when single-step displacements are involved because the evolution operator is diagonal in $k$-space. The Fourier transform $|\tilde{\psi}(k,t)\rangle$ $(k\in(-\pi,\pi))$ is given by
\begin{equation}\label{2-2}
|\tilde{\psi}(k,t)\rangle = \sum_{x} e^{-ikx} |\psi(x,t)\rangle,
\end{equation}
and by the inverse Fourier transform, we have
\begin{equation}\label{2-3}
|\psi(x,t)\rangle= \int_{-\pi}^{\pi} \frac{dk}{2\pi} e^{ikx}|\tilde{\psi}(k,t)\rangle.
\end{equation}
The time evolution of  $|\tilde{\psi}(k,t)\rangle$ after one step of the walk is $|\tilde{\psi}(k,t+1)\rangle = \tilde{U} |\tilde{\psi}(k,t)\rangle$, where $\tilde{U} = R(k)U$ and $R(k) = \left(\matrix{e^{ik}&0\cr0&e^{-ik}}\right)$.

%%%%%%%%%%%%%%%%%%%%%%%%%%%%%%%%%%%%%%%%%%%%%%%%%%%%%%%%%%%%%%%%%%%%%%%%%%%%%%%%%%%%%%%%%%%%%%%%

\section{Entropy of entanglement}
As mentioned above, the conditional shift in evolution operator of a DTQW entangles the coin and position degrees of freedom of the walker. To efficiently make use of entanglement as a physical resource, the amount of entanglement in a given system has to be quantified. Entanglement in a pure bipartite system is quantified using standard measures known as entropy of entanglement corresponds to the von Neumann entropy \cite{43}. It is a functional of the eigenvalues of the reduced density matrix and is given by the formula:
\begin{equation}\label{3-4}
S_{E} = -tr(\rho_{c}\log_{2}\rho_{c}).
\end{equation}
In this equation, $\rho_{c} =tr_{p}(\rho)$ is the reduced density operator obtained from $\rho =(U)^{t}\rho(t=0)(U^{\dag})^{t} $ by tracing out the position degrees of freedom. Since $\rho_{c}$ has two dimension, this quantity is $S_E\in[0,1]$, i.e., $S_{E}=0$ for a product state and $S_{E}=1$ for a maximally entangled state. Note that, in general $tr(\rho_{c})=1$ and $tr (\rho_{c}^{2})\leq 1$. The entropy of entanglement can be obtained after digitalization of
$\rho_{c}$. This operator which acts in $H_{c}$ is represented by the Hermitian matrix as
\begin{equation}\label{3-5}
\rho_{c}=\left(\matrix{\alpha(t)&\beta(t)\cr \beta^{*}(t)&\gamma(t)}\right),
\end{equation}
where
$$\alpha(t)\equiv\sum_{x} |a(x,t)|^{2} = \int_{-\pi}^{\pi} \frac{dk}{2\pi}|\tilde{a}(k,t)|^{2},$$
$$\beta(t)\equiv\sum_{x} a(x,t)b^{*}(x,t) = \int_{-\pi}^{\pi} \frac{dk}{2\pi}\tilde{a}(k,t)\tilde{b}^{*}(k,t),$$
$$\gamma(t)\equiv\sum_{x} |b(x,t)|^{2} = \int_{-\pi}^{\pi} \frac{dk}{2\pi}|\tilde{b}(k,t)|^{2}.$$
Where $*$ denotes the conjugate value. The eigenvalues $r_{1,2}$ of $\rho_{c}$, can be computed as
\begin{equation}\label{3-6}
r_{1,2} = \frac{1}{2}[1\pm\sqrt{1+4(|\beta(t)|^{2}-\alpha(t)\gamma(t))}].
\end{equation}
Therefore, by using Eq. (\ref{3-4}) one can obtain the entropy of entanglement as
\begin{equation}\label{3-7}
S_{E} = -(r_{1}\log{r_{1}}+r_{2}\log{r_{2}}).
\end{equation}
To study asymptotic  entanglement for Hadamard walk, this method has been also used by Abal et al. \cite{35}. They show analytically that for localized particle with every initial coin states, the asymptotic entanglement is 0.872 and when nonlocal initial conditions $|\Psi_{\pm}(0)\rangle=\frac{1}{2}(|-1\rangle\pm|1\rangle)\otimes(|L\rangle+i|R\rangle)$ are considered, the asymptotic entanglement varies smoothly between almost complete entanglement, $S_{E}^{+}\approx 0.979$, and no entanglement (product state), $S_{E}^{-}\approx0.661$.

%%%%%%%%%%%%%%%%%%%%%%%%%%%%%%%%%%%%%%%%%%%%%%%%%%%%%%%%%%%%%%%%%%%%%

\section{ Two-period QWs }
When one quantum coin acts for all the walker position, we face with
a homogeneous walk which the amplitude for moving does not depend on
position but inhomogeneous walk differ from it, in that we allow the
coin operator to depend on the walker position. Now we will define
an inhomogeneous quantum walk in a similar manner to the standard
quantum walk, which the coin operator $C$ to be dependent on $x$. In
the case of walks on the line, $C_{x}$ could be an arbitrary unitary
operator on the two-dimensional coin space. Indeed, there is no need
to restrict to walks in which
 only moves to the nearest neighbours,
more generally one could allow there to be transitions from a given
point to any other point on the line. However for the purposes of
most of our discussion of walks on the line we will focus on the
simplest case of a transitions to nearest neighbors. Let $\{C_{x};
x\in \textbf{Z}\}$ be a sequence of orthogonal matrices with $C_{2s}
= H_{0}$ and $C_{2s+1} = H_{1}$ $(s \in \textbf{Z})$, where
\begin{equation}\label{4-8}
H_{\gamma} = \left(\matrix{\cos\theta_{\gamma} & \sin\theta_{\gamma}\cr \sin\theta_{\gamma} & -\cos\theta_{\gamma}}\right),
\end{equation}
which $\gamma = 0,1$; $\theta_{0}=2\pi \zeta (2s) , \theta_{1}= 2\pi
\zeta (2s+1)$ are free parameters of coin operators which can be
altered to choose the quantum coin operation and $\zeta\in R$ is the
inverse period of the coin operations. With this selection of coin
operators, we face with a two-period quantum walks which are
determined by two orthogonal matrices, and are also self-dual under
the Aubry-Andr\'{e} duality. The inhomogeneous quantum walk is
restricted to the finite interval $[-Q,Q]$, when
$\zeta=\frac{P}{4Q}$ with relatively prime $P$ (odd integer) and
$Q$. So, the limit distribution of the this inhomogeneous quantum
walk,
 divided by any power of the time variable is localized at the origin \cite{55}.
 Limit probability distribution of this two-period QWs computed by Machida et al. \cite{42}.
  In the rest of this work, we shall be concerned with clarifying the asymptotic value of $S_{E}$ for
   both local and nonlocal initial conditions.

%%%%%%%%%%%%%%%%%%%%%%%%%%%%%%%%%%%%%%%%%%%%%%%%%%%%%%%%%%%%%%%%%%%%

\subsection{Asymptotic entanglement from local initial conditions}
Let us first consider in detail the simple case of an initial state localized at the origin with
no CPE as $|\Psi(0)\rangle=|0\rangle\otimes|\psi(0,0)\rangle$,
where $|\psi(0,0)\rangle = a(0,0)|L\rangle + b(0,0)|R\rangle$ is the coin state. Below, we provide
an analytical explanation for the observed values of asymptotic entanglement in the local case.
The time evolution for this initial state in $k$-space regarded as
\begin{equation} \label{4-9}
\left\{\begin{array}{cc} |\tilde{\psi}(k,2t)\rangle = (\tilde{H}_{1}\tilde{H}_{0})^{t} |\tilde{\psi}(k,0)\rangle \,\,\,\, & \mbox{for even times} \\ |\tilde{\psi}(k,2t+1)\rangle = \tilde{H}_{0}(\tilde{H}_{1}\tilde{H}_{0})^{t} |\tilde{\psi}(k,0)\rangle \,\,\,\, & \mbox{for odd times.} \\ \end{array}\right. \end{equation}
We note that $|\widetilde{\psi}(k,0)\rangle = |\psi(0,0)\rangle $
and $\tilde{H}_{\gamma} = R(k)H_{\gamma}$. Two eigenvalues of $\tilde{H}_{1}\tilde{H}_{0}$ are given by
\begin{equation} \label{4-10}
\lambda_{\gamma}(k) = c_{0}c_{1} \cos2k + s_{0}s_{1}+(-1)^{\gamma}i\sqrt{1-(c_{0}c_{1} \cos2k + s_{0}s_{1})^{2}}
 \ for \ (\gamma=0,1),
\end{equation}
where $c_{\gamma} = \cos\theta_{\gamma}$
and $s_{\gamma} = \sin\theta_{\gamma}$. The eigenvectors $|V_{\gamma}(k)\rangle$ corresponding to $\lambda_{\gamma}(k)$ are
\begin{equation}\label{4-11}
\matrix{ |V_{\gamma}(k)\rangle = \frac{1}{\sqrt{N_{\gamma}}}\left(\matrix{u(k) \cr v(k)+(-1)^{\gamma}w(k)}\right)},
\end{equation}
which the elements of this matrix are as follows
$$ u(k) =s_{0}c_{1}e^{2ik}-c_{0}s_{1} $$
$$ v(k) = -ic_{0}c_{1} \sin2k $$
$$ w(k) = i\sqrt{1-(c_{0}c_{1} \cos2k+s_{0}s_{1})^{2}},$$
and $N_{\gamma}$ is the normalization constant.
Using the spectral decomposition for $\tilde{H}_{1}\tilde{H}_{0}$, the Fourier transform $|\widetilde{\psi}(k,2t)\rangle$ is expressed as
\begin{equation}\label{4-12}
|\widetilde{\psi}(k,2t)\rangle = \sum_{\gamma=0}^{1}\lambda_{\gamma}^{t}(k)\langle V_{\gamma}(k)|\widetilde{\psi}(k,0)\rangle|V_{\gamma}(k)\rangle, \end{equation}
and its spinor components are obtain
\begin{equation}\label{4-13}
 \matrix{\widetilde{a}(k,2t)=u( \frac{\lambda_{0}^{t}(k)}{N_{0}}F(k)+\frac{\lambda_{1}^{t}(k)}{N_{1}}G(k)),
\cr \widetilde{b}(k,2t)= v(k)(\frac{\lambda_{0}^{t}(k)}{N_{0}}F(k) +
\frac{\lambda_{1}^{t}(k)}{N_{1}}G(k))+ w(k)(\frac{\lambda_{0}^{t}(k)}{N_{0}}F(k) - \frac{\lambda_{1}^{t}(k)}{N_{1}}G(k)),}
\end{equation}
where
$$F(k)=u^{*}(k)\tilde{a}(k,0)+(v^{*}(k)+w^{*}(k))\tilde{b}(k,0),$$
$$G(k)=u^{*}(k)\tilde{a}(k,0)+(v^{*}(k)-w^{*}(k))\tilde{b}(k,0).$$
Quantum walk after $2t+1$ steps is
\begin{equation}\label{4-14}
|\widetilde{\psi}(k,2t+1)\rangle = \sum_{\gamma=0}^{1}\lambda_{\gamma}^{t}(k)\langle V_{\gamma}(k)|\widetilde{\psi}(k,0)\rangle\tilde{H}_{0}|V_{\gamma}(k)\rangle.
\end{equation}
It is easy to verify that coefficients of $|L\rangle$ and $|R\rangle$ have the forms as
\\ \begin{equation}\label{4-15}
\matrix{\widetilde{a}(k,2t+1)=e^{ik}\{(c_{0}u(k) + s_{0}v(k))(\frac{\lambda_{0}^{t}(k)}{N_{0}}F(k)+\frac{\lambda_{1}^{t}(k)}{N_{1}}G(k))+ s_{0}w(k) (\frac{\lambda_{0}^{t}(k)}{N_{0}}F(k)-\frac{\lambda_{1}^{t}(k)}{N_{1}}G(k))\}
\cr \widetilde{b}(k,2t+1)= e^{-ik}\{(s_{0}u(k)-c_{0}v(k))(\frac{\lambda_{0}^{t}(k)}{N_{0}}F(k) + \frac{\lambda_{1}^{t}(k)}{N_{1}}G(k))- c_{0}w(k)(\frac{\lambda_{0}^{t}(k)}{N_{0}}F(k) - \frac{\lambda_{1}^{t}(k)}{N_{1}}G(k))\}.}
\end{equation}
The relevant quantities for the entropy entanglement are $\alpha(t)$
and $\beta(t)$ which defined in Eq. (\ref{3-5}).
, after $2t$ steps
we obtain
\\\begin{equation}\label{4-16}
\matrix{\alpha(2t)=\int_{-\pi}^{\pi}\frac{dk}{2\pi}\{|u(k)|^{2}(\frac{|F(k)|^{2}}{N_{0}^{2}}+\frac{|G(k)|^{2}}{N_{1}^{2}})\cr+
\frac{|u(k)|^{2}}{N_{0}N_{1}}(\lambda_{0}^{2t}(k)F(k)G^{*}(k)+\lambda_{1}^{2t}(k)F^{*}(k)G(k))\}, \cr\beta(2t)=\int_{-\pi}^{\pi}\frac{dk}{2\pi}\{u(k)v^{*}(k)(\frac{|F(k)|^{2}}{N_{0}^{2}}+\frac{|G(k)|^{2}}{N_{1}^{2}})+
u(k)w^{*}(k)(\frac{|F(k)|^{2}}{N_{0}^{2}}-\frac{|G(k)|^{2}}{N_{1}^{2}})\cr +\frac{u(k)v^{*}(k)}{N_{0}N_{1}}(\lambda_{0}^{2t}(k)F(k)G^{*}(k)+\lambda_{1}^{2t}(k)F^{*}(k)G(k))
-\frac{u(k)w^{*}(k)}{N_{0}N_{1}}(\lambda_{0}^{2t}(k)F(k)G^{*}(k)-\lambda_{1}^{2t}(k)F^{*}(k)G(k))\}.}
\end{equation}
The time dependence of these expressions vanishes in the long time limit by using Riemann-Lebesgue
lemma (see Appendix), and the asymptotic values for even time steps are
\\ \begin{equation}\label{4-17}
\matrix{\overline{\alpha}=\int_{-\pi}^{\pi}\frac{dk}{2\pi}|u(k)|^{2}(\frac{|F(k)|^{2}}{N_{0}^{2}}+\frac{|G(k)|^{2}}{N_{1}^{2}}),
\cr\overline{\beta}=\int_{-\pi}^{\pi}\frac{dk}{2\pi}\{u(k)v^{*}(k)(\frac{|F(k)|^{2}}{N_{0}^{2}}+\frac{|G(k)|^{2}}{N_{1}^{2}})+
u(k)w^{*}(k)(\frac{|F(k)|^{2}}{N_{0}^{2}}-\frac{|G(k)|^{2}}{N_{1}^{2}})\}.}
\end{equation}
\\Where we use overline to indicate that the asymptotic limit has been taken, i.e. $\overline{\rho}_{c}= \lim_{t\rightarrow\infty}\rho_{c}$.
After some algebra and use the Riemann-Lebesgue lemma the asymptotic values
 $\overline{\alpha}$ and $\overline{\beta}$ for odd time steps can be obtained as
\\ \begin{equation}\label{4-18}
\matrix{\overline{\alpha}=\int_{-\pi}^{\pi}\frac{dk}{2\pi}\{(c^{2}_{0}|u(k)|^{2}+c_{0}s_{0}(u(k)v^{*}(k)+u^{*}(k)v(k))+
s^{2}_{0}(|v(k)|^{2}+|w(k)|^{2}))(\frac{|F(k)|^{2}}{N_{0}^{2}}+\frac{|G(k)|^{2}}{N_{1}^{2}})\cr
+(c_{0}s_{0}(u(k)w^{*}(k)+u^{*}(k)w(k))-2s^{2}_{0}v(k)w(k))(\frac{|F(k)|^{2}}{N_{0}^{2}}-\frac{|G(k)|^{2}}{N_{1}^{2}})\}, \cr\overline{\beta}=\int_{-\pi}^{\pi}\frac{dk}{2\pi}e^{2ik}\{(c_{0}s_{0}(|u(k)|^{2}-|v(k)|^{2}-|w(k)|^{2})-c^{2}_{0}u(k)v^{*}(k)+s^{2}_{0}u^{*}(k)v(k)) (\frac{|F(k)|^{2}}{N_{0}^{2}}+\frac{|G(k)|^{2}}{N_{1}^{2}})\cr+(-c^{2}_{0}u(k)w^{*}(k)+s^{2}_{0}u^{*}(k)w(k)+2c_{0}s_{0}v(k)w(k)) (\frac{|F(k)|^{2}}{N_{0}^{2}}-\frac{|G(k)|^{2}}{N_{1}^{2}})\}.}
\end{equation}
In order to further illustrate the effects of bias coin on the
asymptotic entanglement level, the asymptotic entanglement
corresponding to initial condition given by
$|\Psi(0)\rangle=|0\rangle\otimes\frac{|L\rangle+i|R\rangle}{\sqrt{2}}$
is shown in Fig.1 for even times and Fig.2 for odd times,
discretely. The plots in these figures show the entropy of
entanglement as a function of $\theta_{0}$ and $\theta_{1}$ defined
in Eq. (\ref{3-7}). As it is tangible from the form of coin
operator, the period of asymptotic entanglement variations is $\pi$
and the mirroring about the vertical axis indicate that
$S_{E}(\theta_{0},\theta_{1})$ is an even function. Fig.1  (A \ and
\ B) shows that, at the exact value $\theta_{1}=\pi/2$ and all
possible values of $\theta_{0}$ since $\bar{\alpha}=\frac{1}{2}$ and
$\bar{\beta}=\frac{-i}{2}$ so the eigenvalues of $\rho_{c}$ are
$r_{1}=1$ and $r_{2}=0$, thus we face with a completely separable
state and entanglement drops to zero, $S_{E}^{(after \ even \ time \
steps)}(\theta_{0},\frac{\pi}{2})=0$. Moreover for
$\theta_{0}=\pi/2$ and full of $\theta_{1}$ we obtain
$\bar{\alpha}=\frac{1}{2}$ and $\bar{\beta}=\frac{-i}{4}$ thereupon
$r_{1}=\frac{3}{4}$ and $r_{2}=\frac{1}{4}$, in this case we have
$S_{E}^{(after \ even \ time \
steps)}(\frac{\pi}{2},\theta_{1})=0.811278$. In comparison, we find
that, $S_{E}^{(after \ odd \ time \ steps)}$ versus $\theta_{0}$ and
$\theta_{1}$ is symmetric and when these variables take the accurate
quantity $\frac{\pi}{2}$, we have $\bar{\alpha}=\frac{1}{2}$ and
$\bar{\beta}=0$ therefore maximal entanglement shows itself, see
Fig.2 (A \ and \ B). To elaborate this further, we will consider the
steps of evolution of the DTQW using quantum coin operations given
by $H_{0}=\left(\matrix{0&1\cr1&0}\right)$ and
$H_{1}=\left(\matrix{\cos\theta_{1}&\sin\theta_{1}\cr\sin\theta_{1}&-\cos\theta_{1}}\right)$.
After several steps of the DTQW, the state can be written as
\begin{equation}\label{4-19}
\left\{\begin{array}{cc}|\psi(2t)\rangle=\frac{(i)^{t}}{2\sqrt{2}}\{(e^{-it\theta_{1}}+(-1)^{t}e^{it\theta_{1}})
|0\rangle\otimes(|L\rangle+i|R\rangle)\\+(e^{-it\theta_{1}}-(-1)^{t}e^{it\theta_{1}})(|-2\rangle\otimes|L\rangle+i|2\rangle\otimes|R\rangle)\},\\ |\psi(2t+1)\rangle=\frac{(i)^{t}}{2\sqrt{2}}\{(e^{-it\theta_{1}}+(-1)^{t}e^{it\theta_{1}})(i|-1\rangle\otimes|L\rangle+|1\rangle\otimes|R\rangle)\\+
(e^{-it\theta_{1}}-(-1)^{t}e^{it\theta_{1}})(i|1\rangle\otimes|L\rangle+|-1\rangle\otimes|R\rangle)\}.\\ \end{array}\right.
\end{equation}
\begin{figure}\label{fig:Fig1}
\includegraphics{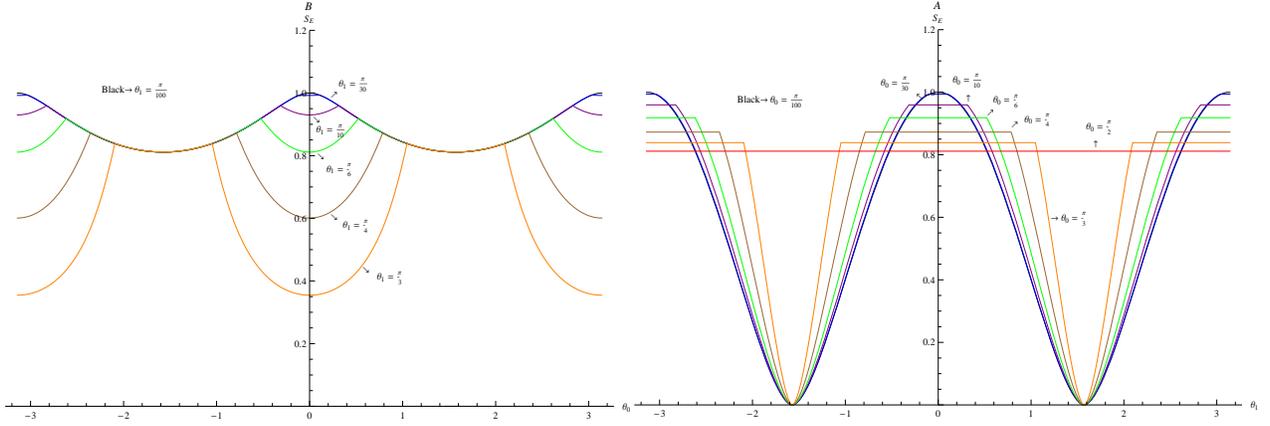}\vspace{7.9cm} \caption{(Colour online) Asymptotic of entanglement $S_{E}$ as a function of
$\theta_{0}$ and $\theta_{1}$ for $\tilde{a}_{0}(k)=i\tilde{b}_{0}(k)=\frac{1}{\sqrt{2}}$ after even time steps. }
\end{figure}
We can observe that, when $H_{0}$ reduces to the Pauli X operator,
then the inverse period of the coin operations corresponds to
$\frac{1}{4\times2}$, so our inhomogeneous quantum random walk is
restricted to the finite interval $[-2,2]$ \cite{55}. The reduced
density operator can be computed from above equation:
\\ $$\rho_{c}(2t)=\left(\matrix{\frac{1}{2}&\frac{-i}{4}(1+(-1)^{t}\cos(2t\theta_{1}))\cr\frac{i}{4}(1+(-1)^{t}\cos(2t\theta_{1}))&\frac{1}{2}}\right) \,\,\,\,and\,\,\,\, \rho_{c}(2t+1)=\left(\matrix{\frac{1}{2}&0\cr0&\frac{1}{2}}\right),$$
\\ after averaging over $t$ in the long time, the matrix element of the reduced density operator is obtained:
$$\overline{\rho}_{c}(after \ enen \ steps)=\left(\matrix{\frac{1}{2}&\frac{-i}{4}\cr\frac{i}{4}&\frac{1}{2}}\right)\,\, \,\,and\,\,\,\, \overline{\rho}_{c}(after \ odd \ steps)=\left(\matrix{\frac{1}{2}&0\cr0&\frac{1}{2}}\right).$$
\\It is easy to check that $S_{E}^{(after \ even \ time \ steps)}(\frac{\pi}{2},\theta_{1})=0.811278$ and $S_{E}^{(after \ odd \ time \ steps)}(\frac{\pi}{2},\theta_{1})=1$, see Figs.1 and 2. When we let $H_{1}$ diminish to coin shift operator and other parameter of coin ($\theta_{0}$) alter, the wave functions of the QW can be written as
\begin{figure}\label{fig:Fig2}
\includegraphics{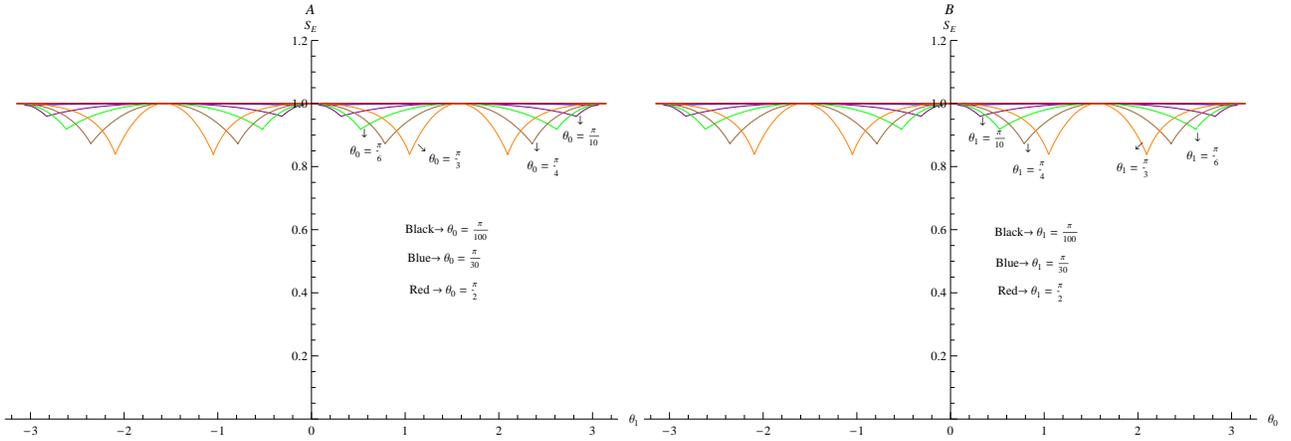}\vspace{7.7cm} \caption{(Colour online) Asymptotic of entanglement $S_{E}$ as a function of
$\theta_{0}$ and $\theta_{1}$ for $\tilde{a}_{0}(k)=i\tilde{b}_{0}(k)=\frac{1}{\sqrt{2}}$ after odd time steps. }
\end{figure}
\begin{equation}\label{4-20}
\left\{\begin{array}{cc}|\psi(2t)\rangle=\frac{(-i)^{t}}{\sqrt{2}}e^{it\theta_{0}}|0\rangle\otimes(|L\rangle+i|R\rangle)\\ |\psi(2t+1)\rangle=\frac{(-i)^{t}}{\sqrt{2}}e^{i(t+1)\theta_{0}}(|-1\rangle\otimes|L\rangle-i|1\rangle\otimes|R\rangle).\\ \end{array}\right.
\end{equation}
It is discernible that particle remain in the origin in all even times as a separable state and in all odd times, it with maximal entanglement exist in the positions $-1$ and $+1$ with coin states $|L\rangle$ and $|R\rangle$, respectively. Hence, the quantum walk is bounded.
\\One of the most remarkable facets of this walk, is behavior of asymptotic CPE for $-\theta_{0}\leq\theta_{1}\leq\theta_{0}$ in all even times. In this interval $S_{E}^{(after \ even \ time \ steps)}$ gets stuck value which is determine by biases of $H_{0}$, this characteristic ascertain that when $\theta_{0}$ attains constant value, conversions of $\theta_{1}$ can not change the level of entanglement. The change of asymptotic CPE is from $\theta_{1}\in(\theta_{0},\pi-\theta_{0})$, in this interval $S_{E}^{(after \ even \ time \ steps)}$ fluctuate between the maximum and minimum possible entanglement levels and its quantity depend on both parameters of coin operator, $\theta_{0}$ and $\theta_{1}$ (see $Fig. \ 1 \ (A)$). Moreover, with increase the value of $\theta_{0}$ from zero to $\frac{\pi}{2}$, maximum quantity of long time limit entanglement reduce. For example we study the behavior of asymptotic entanglement after even time steps for $\theta_{0}=\frac{\pi}{4}$ and we find that $\overline{\alpha}=\frac{1}{2}$ and for $\theta_{1}=\frac{\pm\pi}{4},\frac{\pm3\pi}{4}$, $\overline{\beta}=\frac{i}{2}(1-\sqrt{2})$ but against other values of $\theta_{1}$ exclusion of $\frac{\pm\pi}{2}$ we have
\begin{equation}\label{4-21}
\matrix{\overline{\beta}=\frac{1}{8\pi}\sec^{3}\theta_{1}\{3i\pi\cos\theta_{1}+i\pi\cos3\theta_{1}+
2\cos^{5/2}\cr(\sqrt{3\cos\theta_{1}-\cos3\theta_{1}+4\sqrt{1+\cos2\theta_{1}}\sin\theta_{1}}
\{\ln[\frac{i(1+\cos2\theta_{1}-\sin2\theta_{1}-2\sqrt{1+\cos2\theta_{1}})}
{\sqrt{6\cos^{2}\theta_{1}-2\cos\theta_{1}\cos3\theta_{1}+4\sin2\theta_{1}\sqrt{1+\cos2\theta_{1}}}}]\cr-
\ln[\frac{-i(1+\cos2\theta_{1}-\sin2\theta_{1}-2\sqrt{1+\cos2\theta_{1}})}
{\sqrt{6\cos^{2}\theta_{1}-2\cos\theta_{1}\cos3\theta_{1}+4\sin2\theta_{1}\sqrt{1+\cos2\theta_{1}}}}]\}\cr
+i\sqrt{-3\cos\theta_{1}+\cos3\theta_{1}+4\sqrt{1+\cos2\theta_{1}}\sin\theta_{1}}
\{\ln[\frac{-\sqrt{2}(1+\cos2\theta_{1}-\sin2\theta_{1}+2\sqrt{1+\cos2\theta_{1}})}
{\sqrt{-3\cos^{2}\theta_{1}+\cos\theta_{1}\cos3\theta_{1}+2\sin2\theta_{1}\sqrt{1+\cos2\theta_{1}}}}]\cr-
\ln[\frac{\sqrt{2}(1+\cos2\theta_{1}-\sin2\theta_{1}+2\sqrt{1+\cos2\theta_{1}})}
{\sqrt{-3\cos^{2}\theta_{1}+\cos\theta_{1}\cos3\theta_{1}+2\sin2\theta_{1}\sqrt{1+\cos2\theta_{1}}}}]\})\}.}
\end{equation}
\\when $0\leq\theta_{1}\leq \frac{\pi}{4}$, the exact value of asymptotic CPE is $S_{E}^{(after \ even \ time \ steps)}(\frac{\pi}{4},\theta_{1}\leq \frac{\pi}{4})=0.872429$, this precise value are coincident with that obtained for homogeneous Hadamard walk with local initial condition \cite{35}, from $\frac{\pi}{4}$ to $\pi-\frac{\pi}{4}$ asymptotic CPE variate slowly and achieve zero for correct $\theta_{1}=\frac{\pm\pi}{2}$. While $\theta_{1}$ close to $\pi-\frac{\pi}{4}$, it gradually augment to $0.872429$ when the parameter $\theta_{1}$ attains $\pi-\frac{\pi}{4}$ and the asymptotic entanglement maintain this value to $\theta_{1}=\pi$. In plot $(B)$ presented in $Fig. \ 1$, overlapping graphs endorse our notion that for $\theta_{1}\leq\theta_{0}$ the limiting value of the entanglement is regulated with coin operator $H_{0}$. For example after even time steps we obtain $\overline{\alpha}=\frac{1}{2}$ and for $\theta_{0}=\frac{\pm\pi}{4},\frac{\pm3\pi}{4}$, $\overline{\beta}=\frac{i}{2}(1-\sqrt{2})$ but versus other values of $\theta_{0}$ exclusion of $\frac{\pm\pi}{2}$ we have
\begin{equation}\label{4-22}
\matrix{\overline{\beta}=\frac{1}{4\pi\sqrt{1+\cos2\theta_{0}}\sqrt{-3\cos\theta_{0}+\cos3\theta_{0}+4\sin\theta_{0}\sqrt{1+\cos2\theta_{0}}}}\{i\sec^{3/2}\theta_{0}(-3\cos\theta_{0}+i\cos3\theta_{0}+4\sin\theta_{0}\sqrt{1+\cos2\theta_{0}})\cr
\{\ln[\frac{-(1+\cos2\theta_{0}-\sin2\theta_{0}+2\sqrt{1+\cos2\theta_{0}})}
{\sqrt{-3\cos^{2}\theta_{0}+\cos\theta_{0}\cos3\theta_{0}+2\sin2\theta_{0}\sqrt{1+\cos2\theta_{0}}}}]-
\ln[\frac{(1+\cos2\theta_{0}-\sin2\theta_{0}+2\sqrt{1+\cos2\theta_{0}})}
{\sqrt{-3\cos^{2}\theta_{0}+\cos\theta_{0}\cos3\theta_{0}+2\sin2\theta_{0}\sqrt{1+\cos2\theta_{0}}}}]\}\cr
+i\sec^{3/2}\theta_{0}\sqrt{2\sin^{2}2\theta_{0}-(3\cos\theta_{0}-\cos3\theta_{0})^{2}}\cr
\{\ln[\frac{i(1+\cos2\theta_{0}-\sin2\theta_{0}-2\sqrt{1+\cos2\theta_{0}})}
{\sqrt{3\cos^{2}\theta_{0}-\cos\theta_{0}\cos3\theta_{0}+2\sin2\theta_{0}\sqrt{1+\cos2\theta_{0}}}}]-
\ln[\frac{-i(1+\cos2\theta_{0}-\sin2\theta_{0}-2\sqrt{1+\cos2\theta_{0}})}
{\sqrt{3\cos^{2}\theta_{0}-\cos\theta_{0}\cos3\theta_{0}+2\sin2\theta_{0}\sqrt{1+\cos2\theta_{0}}}}]\}\cr
+2\sqrt{2}i\sqrt{1+\cos2\theta_{0}}\sqrt{-3\cos\theta_{0}+\cos3\theta_{0}+4\sin\theta_{0}\sqrt{1+\cos2\theta_{1}}}\tan^{2}\theta_{0}\}.}
\end{equation}

\subsection{Asymptotic entanglement from nonlocal initial conditions}
Asymptotic entanglement levels for nonlocal initial conditions are reported for the first time in the context of the homogeneous Hadamard walk on the line. In order to show that the asymptotic entanglement level is strongly dependent on whether the initial condition is localized or delocalized in position space. In this section we apply our inhomogeneous walk to consider in detail the case of initial conditions in the position subspace $\mathbf{H}_{p}$ spanned by $|\pm1\rangle$, as
\begin{equation}\label{4-23}
|\Psi(0)\rangle=\frac{|-1\rangle+|1\rangle}{\sqrt{2}}\otimes\frac{|L\rangle+i|R\rangle}{\sqrt{2}}.
\end{equation}
Whit the projection on k-space this non localized state is simply $|\tilde{\psi}(k,0)\rangle=\cos(k)(|L\rangle+i|R\rangle)$. The dependence on the initial conditions is contained in the coin operator. The time evolution of the motion in phase space for the initial states defined in the previous equation can be expressed by
\begin{equation}\label{4-24}
\left\{\begin{array}{cc} |\tilde{\psi}(k,2t)\rangle = (\tilde{H}_{0}\tilde{H}_{1})^{t} |\tilde{\psi}(k,0)\rangle \,\,\,\, & \mbox{for even times} \\ |\tilde{\psi}(k,2t+1)\rangle = \tilde{H}_{1}(\tilde{H}_{0}\tilde{H}_{1})^{t} |\tilde{\psi}(k,0)\rangle \,\,\,\, & \mbox{for odd times.} \\ \end{array}\right.     \end{equation}
\begin{figure}\label{fig:Fig3}
\includegraphics{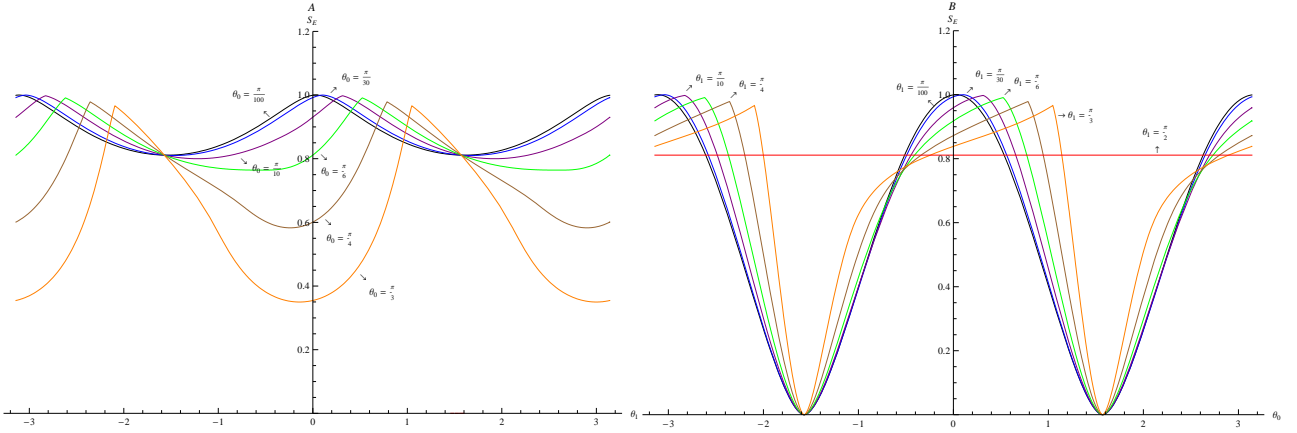}\vspace{5.7cm} \caption{(Colour online) Asymptotic of entanglement $S_{E}$ as a function of
$\theta_{0}$ and $\theta_{1}$ for $\tilde{a}_{0}(k)=i\tilde{b}_{0}(k)=\cos(k)$ in even times. }
\end{figure}
One can check that the eigenvalues of $\tilde{H}_{0}\tilde{H}_{1}$ are equivalent with Eq. (\ref{4-10}), exactly, and the normalized eigenvectors are of the form Eq. (\ref{4-11}) with components
$$ u'(k)=s_{1}c_{0}e^{2ik}-c_{1}s_{0} $$
$$ v(k)=-ic_{0}c_{1}\sin2k $$
$$ w(k)=i\sqrt{1-(c_{0}c_{1}\cos2k+s_{0}s_{1})^{2}}. $$
As in previous subsection relate in details, the matrix elements of the reduced density operator after even times is obtained as
\\ \begin{equation}\label{4-25}
\matrix{\overline{\alpha}=\int_{-\pi}^{\pi}\frac{dk}{2\pi}|u'(k)|^{2}(\frac{|F(k)|^{2}}{N_{0}^{2}}+\frac{|G(k)|^{2}}{N_{1}^{2}}),
\cr\overline{\beta}=\int_{-\pi}^{\pi}\frac{dk}{2\pi}\{u'(k)v^{*}(k)(\frac{|F(k)|^{2}}{N_{0}^{2}}+\frac{|G(k)|^{2}}{N_{1}^{2}})+
u'(k)w^{*}(k)(\frac{|F(k)|^{2}}{N_{0}^{2}}-\frac{|G(k)|^{2}}{N_{1}^{2}})\}.}
\end{equation}
\\Subsequently, after odd times the relevant quantities for the entropy of entanglement are
\\ \begin{equation}\label{4-26}
\matrix{\overline{\alpha}=\int_{-\pi}^{\pi}\frac{dk}{2\pi}\{(c^{2}_{1}|u'(k)|^{2}+c_{1}s_{1}(u'(k)v^{*}(k)+u'^{*}(k)v(k))+
s^{2}_{1}(|v(k)|^{2}+|w(k)|^{2}))(\frac{|F(k)|^{2}}{N_{0}^{2}}+\frac{|G(k)|^{2}}{N_{1}^{2}})\cr
+(c_{1}s_{1}(u'(k)w^{*}(k)+u'^{*}(k)w(k))-2s^{2}_{1}v(k)w(k))(\frac{|F(k)|^{2}}{N_{0}^{2}}-\frac{|G(k)|^{2}}{N_{1}^{2}})\}, \cr\overline{\beta}=\int_{-\pi}^{\pi}\frac{dk}{2\pi}e^{2ik}\{(c_{1}s_{1}(|u'(k)|^{2}-|v(k)|^{2}-|w(k)|^{2})-c^{2}_{1}u'(k)v^{*}(k)+s^{2}_{1}u'^{*}(k)v(k)) (\frac{|F(k)|^{2}}{N_{0}^{2}}+\frac{|G(k)|^{2}}{N_{1}^{2}})\cr+(-c^{2}_{1}u'(k)w^{*}(k)+s^{2}_{1}u'^{*}(k)w(k)+2c_{1}s_{1}v(k)w(k)) (\frac{|F(k)|^{2}}{N_{0}^{2}}-\frac{|G(k)|^{2}}{N_{1}^{2}})\}.}
\end{equation}
\begin{figure}\label{fig:Fig4}
\includegraphics{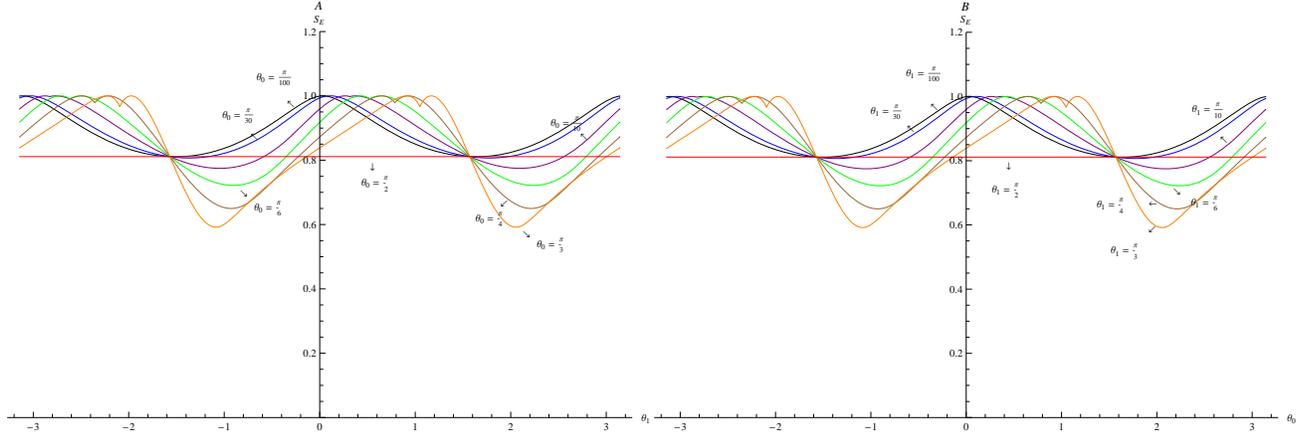}\vspace{6.7cm} \caption{(Colour online) Asymptotic of entanglement $S_{E}$ as a function of
$\theta_{0}$ and $\theta_{1}$ for $\tilde{a}_{0}(k)=i\tilde{b}_{0}(k)=\cos(k)$ in odd times. }
\end{figure}
From these two above equations, one readily obtains the independent elements of $\overline{\rho}_{c}$ for this kind of initial conditions. $Fig. \ 3 \ (A \ and \ B)$ shows the entropy of entanglement as a function of $\theta_{0}$ and $\theta_{1}$ after even time steps and $Fig. \ 4 \ (A \ and \ B)$ shows it after odd time steps. In comparison with local initial condition, the first aspect of these graphs to highlight is that, asymptotic entanglement in odd times versus both variations of coin operator is symmetric and this distinguishing quality is free from type of initial conditions. The second aspect is that, $S_{E}^{(after \ even \ time \ steps)}$ vanishes which $\theta_{0}$ tends to become $\frac{\pi}{2}$, also there is $S_{E}^{(after \ even \ time \ steps)}(\theta_{0},\frac{\pi}{2})=0.811278$, (see $Fig. \ 3$).
The origin of all the above facts can be clarified by using the analytical expressions. Let us begin by $H_{1}=\left(\matrix{0&1\cr1&0}\right)$ and $H_{0}=\left(\matrix{\cos\theta_{0}&\sin\theta_{0}\cr\sin\theta_{0}&-\cos\theta_{0}}\right)$, the wave functions of QW are
\begin{equation}\label{4-27}
\left\{\begin{array}{cc}|\psi(2t)\rangle=\frac{(i)^{t}}{4}\{(e^{-it\theta_{0}}+(-1)^{t}e^{it\theta_{0}})
(|1\rangle\otimes|L\rangle+i|-1\rangle\otimes|R\rangle)\\+(e^{-it\theta_{0}}-(-1)^{t}e^{it\theta_{0}})(|-3\rangle\otimes|L\rangle+i|3\rangle\otimes|R\rangle)\}+
\frac{(i)^{t}}{2}e^{-it\theta_{0}}(|-1\rangle\otimes|L\rangle+i|1\rangle\otimes|R\rangle),\\ |\psi(2t+1)\rangle=\frac{(i)^{t}}{4}\{(e^{-it\theta_{0}}+(-1)^{t}e^{it\theta_{0}})(i|-2\rangle\otimes|L\rangle+|2\rangle\otimes|R\rangle)\\+
(e^{-it\theta_{0}}-(-1)^{t}e^{it\theta_{0}})(i|2\rangle\otimes|L\rangle+|-2\rangle\otimes|R\rangle)\}+
\frac{(i)^{t}}{2}e^{-it\theta_{0}}|0\rangle\otimes(i|L\rangle+|R\rangle).\\ \end{array}\right.
\end{equation}
Density operator after some manipulation, can be expressed as
$$\rho_{c}(2t)=\left(\matrix{\frac{1}{2}&\frac{-i}{4}(1+(i)^{t}e^{2it\theta_{0}})\cr\frac{i}{4}(1+(-i)^{t}e^{-2it\theta_{0}})&\frac{1}{2}}\right) \,\,\,\,and\,\,\,\, \rho_{c}(2t+1)=\left(\matrix{\frac{1}{2}&\frac{i}{4}\cr\frac{-i}{4}&\frac{1}{2}}\right),$$
in the long-time limit, the contribution of the time dependent terms in these $\rho_{c}$s vanishes and we have
$$\overline{\rho}_{c}(after \ enen \ steps)=\left(\matrix{\frac{1}{2}&\frac{-i}{4}\cr\frac{i}{4}&\frac{1}{2}}\right)\,\, \,\,and\,\,\,\, \overline{\rho}_{c}(after \ odd \ steps)=\left(\matrix{\frac{1}{2}&\frac{i}{4}\cr\frac{-i}{4}&\frac{1}{2}}\right).$$
From Eq. (\ref{3-6}), the exact eigenvalues of the both operators are $r_{1}=\frac{3}{4}$ and $r_{2}=\frac{1}{4}$, these eigenvalues yield the asymptotic value for the entropy of entanglement
$$S_{E}=-(\frac{3}{4}\log_{2}(\frac{3}{4})+\frac{1}{4}\log_{2}(\frac{1}{4}))= 0.811278.$$
If $H_{0}$ replace by the shift coin operator, then the position states are
\begin{equation}\label{4-28}
\left\{\begin{array}{cc}|\psi(2t)\rangle=\frac{(-i)^{t}}{2}e^{it\theta_{1}}(|-1\rangle+|1\rangle)\otimes(|L\rangle+i|R\rangle)\\ |\psi(2t+1)\rangle=\frac{(-i)^{t}}{\sqrt{2}}e^{i(t+1)\theta_{1}}|0\rangle\otimes(|L\rangle+i|R\rangle).\\ \end{array}\right.
\end{equation}
Walk is bounded in this cases, overtly, the reduce density operators are given by
$$\rho_{c}(2t)=\left(\matrix{\frac{1}{2}&\frac{i}{2}\cr\frac{-i}{2}&\frac{1}{2}}\right) \,\,\,\,and\,\,\,\, \rho_{c}(2t+1)=\left(\matrix{\frac{1}{2}&\frac{i}{4}\cr\frac{-i}{4}&\frac{1}{2}}\right),$$
we observe that entropy of entanglement is zero and $0.811278$ in the even and odd times respectively.

\section{Conclusion and discuss}
We have investigated an inhomogeneous walk with two-period, to be used in quantum random walk in one dimension latices. It determined by two orthogonal matrices and included two free parameters that together provided many conditions under which a measurement performed on the coin state yielded the value of entanglement on the resulting position quantum state. We have studied the problem analytically for all values of two free parameters of coin with diverse initial conditions for large number of QW steps. To summarise the results for this walk, we find the various behaviours of the entanglement are governed as follows. We demonstrated that \textbf{how} the asymptotic value depend on coin parameters and initial condition. Moreover it was different for odd or even positions and when on the coin parameters take the precise value $\frac{\pi}{2}$ walk altered to a bounded motion. Two striking characteristics of this walk was that, when motion began with local initial condition, walk regulated by the first coin operator that act on the state of particle in all even tims and $S_{E}^{(after \ odd \ time \ steps)}$ was symmetric ratio $\theta_{0}$ and $\theta_{1}$ for all type of initial conditions.

\vspace{1.5cm}
\setcounter{section}{0}
\setcounter{equation}{0}
\renewcommand{\theequation}{A-\roman{equation}}
{\Large{\textbf{Appendix}}}
{ \begin{center}
\textbf{The Riemann-Lebesgue lemma}
\end{center}}
This little note is devoted to a proof of the Riemann-Lebesgue lemma. We use the following notation for the $n$th Fourier coefficient of a $2\pi$-periodic
function $f$:
$$f(n)=\int_{-\pi}^{\pi}\frac{dk}{2\pi}\tilde{f}(k)e^{ink}. $$
Lemma: Assume that $\tilde{f}(k)$ is $2\pi$-periodic, bounded and integrable. Then $f(n)\rightarrow 0$ when $n\rightarrow \pm\infty$.
\\Proof: We shall prove this only for real valued functions. If $\tilde{f}(k)$ is complex valued, the result will follow from the result applied to the real and imaginary parts of $\tilde{f}(k)$ separately. First, we prove the result for an extremely special case: Namely, a single step, which is a function of the form
\[\tilde{s}(k)=\left\{\begin{array}{cc} 1 & \ a+2k\pi\leq k \leq b+2k\pi, \ \ k \in \textbf{Z}  \\  0 &  \ otherwise \end{array}\right.\]
where $a<b$ and $b-a<2\pi$. Then
$$s(n)=\int_{a}^{b}\frac{dk}{2\pi}e^{ink}=\frac{e^{inb}-e^{ina}}{2\pi in}\rightarrow 0  \ \ \ as \ n\rightarrow\pm\infty,$$
since the numerator is bounded and the denominator goes to infinity. Second, since any step function is a linear combination of a finite number of single steps, the same result holds for step functions.
\\Finally, now assume that $\tilde{f}(k)$ is integrable, and pick any $\varepsilon>0$. It follows - practically direct from the definition of integrability - that there exists a step function $\tilde{s}(k)$ with
$$\int_{-\pi}^{\pi}\frac{dk}{2\pi}|\tilde{f}(k)-\tilde{s}(k)|<\varepsilon.$$
From this we get
$$|f(n)-s(n)|=|\int_{-\pi}^{\pi}\frac{dk}{2\pi}(\tilde{f}(k)-\tilde{s}(k))e^{ink}|\leq\int_{-\pi}^{\pi}\frac{dk}{2\pi}|\tilde{f}(k)-\tilde{s}(k)|<\varepsilon,$$
as well. We have shown that $s(n)\rightarrow 0$, so there is some $N$ so that $|n|\geq N$ implies $s(n)<\varepsilon$. Whenever $|n|\geq N$, then
$$|f(n)|\leq|f(n)-s(n)|+|s(n)|<\varepsilon+\varepsilon=2\varepsilon,$$
which finishes the proof.
\\Notice that the Riemann-Lebesgue lemma says nothing about how fast $f(n)$ goes to zero. With just a bit more of a regularity assumption on $\tilde{f}(k)$, we can show that $f(n)$ behaves roughly like $1/n$ or better. This is easy if $\tilde{f}(k)$ is continuous and piecewise smooth, as is seen from the identity
$f'(n)=i \ n f(n)$, which arises from partial integration. Applying the Riemann–Lebesgue lemma to $\tilde{f}'(k)$ we conclude that $f(n)$ is $1/n$ times something that goes to zero, so $f(n)\rightarrow 0$ faster than $1/n$, \cite{46}.

\end{document}